# Evidence for anomalous structural behavior in polycrystalline $CaFe_2As_2$

S. K. Mishra, R. Mittal, P. S. R. Krishna, P. U. Sastry and S. L. Chaplot

*Solid State Physics Division, Bhabha Atomic Research Centre, Trombay, Mumbai 400 085, India.*

P D Babu

*UGC-DAE Consortium for Scientific Research Mumbai Centre, BARC, Mumbai-400085, India*

S. Matsuishi

*Materials Research Center for Element Strategy, Tokyo Institute of Technology, Nagatsuta-cho, Midori-ku, Yokohama 226-8503, Japan*

H. Hosono

*Frontier Research Center, Tokyo Institute of Technology, Nagatsuta-cho, Midori-ku, Yokohama 226-8503, Japan*

## Abstract

The structural properties of the $CaFe_2As_2$ have been investigated by x-ray and neutron powder diffraction techniques as a function of temperature. Unambiguous experimental evidence is shown for coexistence of tetragonal and orthorhombic phases below 170 K in contrast to existing literature. Detailed Rietveld analyses of thermo-diffractograms show that the sample does not transform completely in to the orthorhombic phase at the lowest temperature of 5 K even though it is the majority phase. We have found that the unit cell volume of the orthorhombic phase is higher compared to that of the tetragonal phase for all the temperatures. X-ray data on $CaFe_2As_2$ shows anomalous ($a_t$) lattice parameter contraction with increasing temperature and phase co-existence behavior below 170 K unlike other members of the 122 family of compounds like $SrFe_2As_2$ and $EuFe_2As_2$. Temperature dependent magnetization of polycrystalline $CaFe_2As_2$ sample show weak anomalies below 170 K. This behavior of the polycrystalline sample is in contrast to that of a single crystal reported earlier.

## Introduction

The discovery of Fe-based superconductors by Hosono group [1], opened a new frontier of superconductivity [2–6]. The key player in these materials is the element iron (Fe), which was earlier known to be an element for destabilization of superconductivity. Compositionally, various types of Fe-based superconductors have been reported in literature. But, crystallo-graphically, their parent materials have been classified into five types [4]. A wide variety of experimental and theoretical techniques have been used to understand physics of 1111 and 122 types of Fe-based compounds [3–33]. At ambient pressure, they possess similar sequence of structural phase transitions with decreasing temperature (tetragonal to orthorhombic). But, our recent high pressure synchrotron study on these compounds reveals a different phase transition sequence [20]. The parent materials of both 1111 and 122 families do not exhibit superconductivity at ambient pressure, but become superconductors by application of chemical or external pressure.

$CaFe_2As_2$ has been investigated extensively by different techniques. At ambient conditions, it crystallizes in to tetragonal symmetry with space group *I4/mmm*. On cooling, it transforms to orthorhombic structure at 170 K which is accompanied by a magnetic transition. Using single crystal x-ray diffraction technique, Ni *et al* [12] have shown that the tetragonal to the orthorhombic phase transition is of first order type with hysteresis and coexistence over a temperature range of 2 to 3 K. Physical properties viz., specific heat, electrical resistance and magnetic susceptibility exhibit an anomaly at structural and magnetic phase transition temperature (170 K) [10, 12]. Similarly, measurements on structural and physical properties of $BaFe_2As_2$, $SrFe_2As_2$, and $EuFe_2As_2$ compounds suggest that the tetragonal to orthorhombic phase transition in these materials occurs at about 140 K, 205 K and 210 K [13, 14, 30] respectively. Ran et al. [33] have shown that the tetragonal to orthorhombic phase transition can be suppressed by manipulating the annealing temperature. Recent Transmission Electron Microscopy (TEM) studies [19] reveal that the $CaFe_2As_2$ has more complex microstructure properties than $SrFe_2As_2$.

We have earlier carried out high temperature diffraction measurements [34] on 122 compounds viz., $CaFe_2As_2$ and $BaFe_2As_2$. We found that the lattice parameters of $BaFe_2As_2$ monotonically increase with increasing temperature in the tetragonal phase. In contrast, for $CaFe_2As_2$, though the c-parameter monotonically increases with increasing temperature, the a-parameter decreases with increasing temperature up to 600 K.

In the present study, we have extended our investigations to low temperatures. We have carried out systematic temperature-dependent X-ray and neutron powder diffraction measurements. Temperature dependence of structure evolution, and structural parameters were examined in detail. Neutron diffraction offers certain advantages over X-rays especially in the accurate determination of the atomic positions and thermal parameters. Interestingly, in contrast to earlier reports in literature, we noticed the coexistence of tetragonal phase with orthorhombic phase over a wide range of temperatures for polycrystalline $CaFe_2As_2$. We also carried out diffraction measurements up to 600 K on $SrFe_2As_2$ and $EuFe_2As_2$ that agree with published results below 300 K. Usual thermal expansion behaviour is observed in all the lattice parameters of $CaFe_2As_2$, $SrFe_2As_2$ and $EuFe_2As_2$ compound except the a- lattice parameter of $CaFe_2As_2$. Temperature dependent magnetization of polycrystalline $CaFe_2As_2$ sample show weak anomalies below 170 K. This behavior of the polycrystalline sample is in contrast to that of a single crystal reported earlier.

## Experimental

$CaFe_2As_2$ samples were prepared by a solid state reaction method as in our earlier work [21]. CaAs and $Fe_2As$ precursors were used as starting materials to achieve better homogeneity. CaAs compounds were synthesized by heating Ca shots (99.99 wt%) with As powder (99.9999 wt%) at 650 ◦C for 10 h in an evacuated silica tube. $Fe_2As$ was synthesized from powders of mixed elements at 800 ◦C for 10 h (Fe, 99.9 wt%). These products were then mixed by agate mortar in stoichiometric ratios, pressed, and heated in evacuated silica tubes at 900 ◦C for 10 h to obtain sintered pellets. All the starting material preparation procedures were carried out in an argon-filled glove box ($O_2$, $H_2O$ < 1 ppm). The X-ray and neutron powder diffraction measurements for $CaFe_2As_2$ at ambient temperature confirmed the sample to be in a single-phase. All the reflections can be indexed with tetragonal symmetry with space group *I4/mmm* consistent with the published reports [10, 12]. The neutron powder diffraction is performed as a function of temperature at the Dhruva reactor at Bhabha Atomic Research Centre, Trombay, India. The structural refinements were performed using the Rietveld refinement program FULLPROF [35]. Magnetization measurements were performed using a 9 Tesla PPMS-VSM (make Quantum Design) at UGC-DAE CSR, Mumbai.

## Result and Discussion

Figure 1 (a) shows both zero field cooled (ZFC) and field cooled (FC) magnetization (M) curves of polycrystalline $CaFe_2As_2$ measured in a magnetic field of $H$ = 1 kOe. As the temperature decreases, the magnetization of the polycrystalline sample increases below 170 K show a maximum at 150 K. On further lowering the temperature magnetization starts decreasing and shows a minima around 100 K. Below this temperature, the magnetization shows a Curie-Weiss-like behavior, increasing rapidly with decrease in temperature. Later observation is in contrast to the magnetization data of the $CaFe_2As_2$ single crystal known to show a strong anomaly at ~ 165 K in the *H//ab* plane and H//c only. The susceptibility transformation was attributed to the occurrence of tetragonal-to-orthorhombic and magnetic phase transitions. Inset shows magnified view of anomaly in magnetization with field cooled (FC) during cooling (FCC) and warming (FCW) cycles. The onset of transition below 170 K manifests itself in the form of magnetic hysteresis as seen clearly between the cooling and heating magnetization curves. . The smearing of anomaly in polycrystalline sample is very similar to the behavior observed in the $Ba_{1-x}K_x$ $Fe_2As_2$, and $(Ca_{1-x}Na_x)$ $Fe_2As_2$ systems [36, 37]. The variation of R (T)/ R(210 K) with temperature also shows a change in the slope around the same temperature. Also a finite resistivity seen even at very low temperatures (up to 3K) clearly rules out superconductivity in this compound at ambient pressure. We measured magnetization versus magnetic field isotherms at different temperatures shown in lower panel of figure 1 (b). Presence of a small hysteresis loop in magnetization vs magnetic field data at 2 K is a signature for weak ferromagnetism. However, we have not observed hysteresis loops at other temperatures. To investigate the phase transition in polycrystalline $CaFe_2As_2$ sample, we carried out neutron, and x-ray diffraction measurements as a function of temperature.

The structural and magnetic phase transitions in the parent $AFe_2As_2$ (122 family) compounds are concomitant in temperature. However, the nature of the phase transition varies in different compounds. For $BaFe_2As_2$, it has been argued by some groups that the structural and magnetic transitions are second-order by showing that the transition is continuous, while first-order behavior such as a large hysteresis has been observed by other groups [38, 39] and found to be dependent on sample preparation method.

Figs. 2 (a) and (b) depict the evolution of the (1 1 2) and (0 0 8) profiles (indexed with tetragonal phase) for x-ray data as a function of temperature. Below 175 K, (1 1 2) profile first broaden up to 95 K and then splits in to two peaks. This is unambiguous signature for stabilization of the orthorhombic phase. It is also seen that the intensity of (0 0 8) ( $Q \approx 4.35$ Å$^{-1}$ ) of tetragonal phase decreases on lowering temperature and shifts towards higher Q values. This is accompanied by the appearance of an additional peak around $Q \approx 4.32$ Å$^{-1}$ ($\leq$ 125 K) and its intensity increases on lowering the temperature. It is evident from the Fig. 2(c), that powder neutron diffraction patterns also show dramatic changes with temperature especially in terms of dissimilar broadening, intensity and splitting of various peaks. Similar to powder x-ray diffraction, we have also noticed the appearance of an additional reflection around $Q \approx 4.32$ Å$^{-1}$ below 130 K, which corresponds to (0 0 8) reflection of tetragonal phase. Further, the intensity of this peak increases gradually with decrease of temperature. It is noteworthy that this peak could not be assigned to either of the orthorhombic or tetragonal phases. Detailed features of the phase transition and its behavior are examined by analyzing the powder diffraction data at various temperatures and the results are discussed below. At low temperatures we also observed additional reflections in neutron patterns due to the magnetic ordering in the sample.

Rietveld refinement of the powder diffraction data shows that the temperature dependent x-ray and neutron diffraction patterns could be indexed using the tetragonal symmetry from 300 K to 175 K. The splitting of (1 1 2) peak of tetragonal phase in powder diffraction patterns indicates a structural phase transition from tetragonal to orthorhombic phase at 170 K which is well documented in literature. Hence, we made an attempt to refine the data using orthorhombic structure with space group *Fmmm* below this temperature. However, all the peaks could not be accounted in refinement with only one phase, i.e. either the orthorhombic or the tetragonal phase. Instead, a two-phase refinement with both the orthorhombic and tetragonal space groups resulted in a successful fit and all the observed diffraction peaks could be clearly indexed. Rietveld refinement (Fig.3) of neutron and X-ray powder diffraction data (in terms of Q ($Å^{-1}$), which enables an easy comparison between them) shows unambiguous evidence for phase coexistence. This indicates that structural phase transition from the orthorhombic to the tetragonal phase is of first order in nature. Rietveld refinements employing this two phase model are satisfactory for all the diffraction patterns up to the lowest temperature measured by us for X-ray (12 K) and neutron (6 K) data. The percentage of the orthorhombic phase fraction was found to increase on lowering the temperature. Rietveld analysis on the neutron data has also been carried out for determining the magnetic structure in the orthorhombic phase at 6 K. We found that the Fe moment is 0.83(3) $\mu_B$ along the longer a-axis with the magnetic propagation vector along (101). The obtained moment and propagation vector are consistent with literature [11, 30]. In all Rietveld refinements data over full angular range has been used, although only a limited range is shown for clarity.

Figure 4 shows evolution of structural parameters of tetragonal phase and orthorhombic phase as a function of temperature. It is evident from this figure that the a- lattice parameter ($a_t$) of tetragonal phase increases with decreasing temperature in the entire temperature range. On the other hand, the c-parameter ($c_t$) of tetragonal phase decreases with decreasing temperature up to 50 K below which it becomes nearly independent of temperature. For the orthorhombic phase, the lattice parameter $a_o$ increases and $b_o$ decreases with decreasing temperature whereas $c_o$ is nearly constant up to 6 (12) K in neutron (x-ray) measurements. It is clear that the rate of change of lattice parameters in the temperature range 120 −170 K is higher as compared with that in the remaining temperatures. It is also noticed that the unit cell volume of tetragonal phase (Fig 5 (a)) decreases with decreasing temperature down to 90 K below which it increases. In contrast, unit cell volume of the orthorhombic phase is nearly constant with the temperature. Also, it is important to note that the orthorhombic phase has higher unit cell volume compared to that of tetragonal phase at all the temperatures. The phase fraction obtained as a function of temperature for the tetragonal phase is shown in Fig. 5(b). On increasing the temperature, the fraction for tetragonal phase increases gradually up to 100 K with an abrupt increase above 100 K. Rietveld refinement of the diffraction data (Fig.2 ) clearly reveals that for T= 6 K, the majority phase is orthorhombic (95%). In other words, the sample does not transform completely even at the lowest temperature.

In order to further emphasize the anomalous behavior of $CaFe_2As_2$, we have also examined the temperature variation of lattice parameters of other members of the family, namely, $SrFe_2As_2$ and $EuFe_2As_2$. We found that $a_t$ of the tetragonal phase of both the compounds decreases with decreasing temperature in entire temperature range (Fig. 6), in contrast to $CaFe_2As_2$. The $c_t$ of tetragonal and $c_o$ of orthorhombic phases for both the compounds decrease while cooling with an anomaly at respective transition temperatures. The behavior of $a_o$ and $b_o$ lattice parameters of the orthorhombic phase is found to be consistent with literature. These results do not indicate coexistence of phases over a wide range of temperature in $SrFe_2As_2$ and $EuFe_2As_2$ compounds unlike in the $caFe_2As_2$ compound.

Figure 7 depicts the anisotropic thermal expansivities (relative to the values at T=0 K) i.e. $A_T = (a_t(T) - a_t(0))/a_t(0)$ and $C_t = (c_t(T) - c_t(0))/c_t(0)$ with temperature for all the compounds. It is clear from this figure that $A_T$ of $CaFe_2As_2$ contracts with increasing temperature, whereas $C_T$ expands. The rate of contraction in $A_T$ is lower than the rate of expansion in $C_T$. Other compounds show normal thermal expansion behaviour along both a and c axes. It may be noted that the in-plane expansion (i.e. along a- axis) is found to be the smallest for $EuFe_2As_2$ and the highest for $BaFe_2As_2$. However, the rate of change of thermal expansivities out-of-plane (i.e.along c-axis) is higher as we go from compounds of Ba, Sr, Eu and Ca, respectively.

For a deeper understanding of the behavior of tetragonal to orthorhombic phase transition from microscopic point of view, we identify the symmetry mode ($\Gamma_1^+$), which characterizes this transformation. The program AMPLIMODES [40] was used to calculate the magnitude of the ($\Gamma_1^+$) distortion mode of *Fmmm* and also the degree of lattice distortion (S), which is the spontaneous strain (sum of the squared eigenvalues of the strain tensor divided by 3). The results are shown in Fig. 8. The (($\Gamma_1^+$)) mode corresponds to a symmetric breathing mode bringing the change in z coordinate of the As atoms only. This could be treated as an order parameter for the orthorhombic to the tetragonal phase.

For $CaFe_2As_2$, the lattice strain increases sharply with decrease in temperature up to about 140 K and becomes invariant below this temperature (Fig. 8(a)). For $SrFe_2As_2$ and $EuFe_2As_2$, strain increases with decreasing temperature but at a different rate. The distor-tion of amplitude mode also exhibits anomalous behavior for $CaFe_2As_2$. The orthorhombic splitting $[(a_o - b_o)/(a_o + b_o)]$ could be treated as an additional order parameter for orthorhombic to tetragonal phase transition. Figure 7(c) depicts the temperature dependence of the orthorhombic splitting $[(a_o - b_o)/(a_o + b_o)]$ for $CaFe_2As_2$, $SrFe_2As_2$ and $EuFe_2As_2$ respectively. We found the orthorhombic splitting abruptly increases with decreasing temperature up to 150 K and then the rate of increment is slower. Thus, our study shows unambiguous evidence for the presence of the tetragonal phase coexistence with the orthorhombic phase over a wide range of temperatures and found to be in contrast from literature [33].

To understand the origin of the phase transition in these materials, extensive theoretical studies have been performed by various researchers [8, 28, 31, 32]. Theoretical calculations on $CaFe_2As_2$ have shown that the orthorhombic phase has higher volume compared to the tetragonal phase. This is in agreement with our diffraction data. Further, the total energy difference between the orthorhombic and the tetragonal phase is $\approx$ 1 meV/degree of freedom and such a small difference in total energy suggests that energetically both phases could be stabilized at finite temperatures, thus giving the possibility for the coexistence of both phases. Our experimental observations are consistent with this picture.

## Conclusion

In summary, we have provided unambiguous experimental evidence for the phase coexistence in polycrystalline $CaFe_2As_2$ in contrast with other 122 compounds studied. Detailed analysis of temperature dependence of powder x-ray and neutron data shows that both orthorhombic and tetragonal phases coexist down to the lowest temperature studied. The phase fraction of the orthorhombic phase has been found to increase on decreasing the temperature. The orthorhombic phase has higher unit cell volume compared to that of the tetragonal phase. We have found that the magnetic structure determined by us for the orthorhombic phase is consistent with the literature. We observed weak anomaly in magnetization around 100 K in polycrystalline sample indicating growth of magnetic phase below this temperature.

### Acknowledgments

We would like to thank Mr. V. B. Jayakrishnan, Solid State Physics Division, Bhabha Atomic Research Centre, Mumbai, India for x-ray measurements

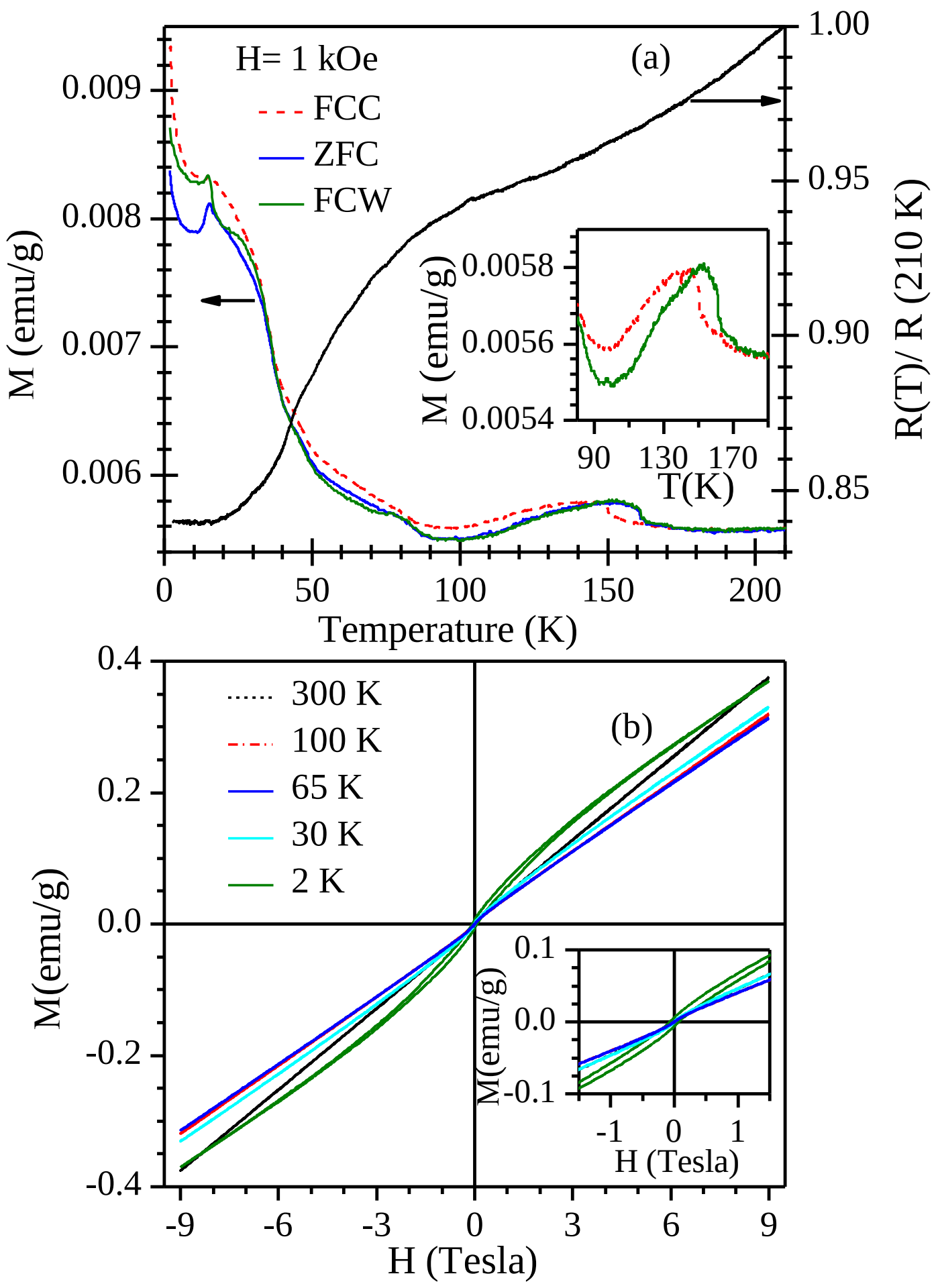

**FIG. 1:** (colour online) (a) Variation of magnetization with zero field cooled (ZFC), field cooled (FC) and normalized resistance (R(T)/R(210 K)) as a function of temperature for $CaFe_2As_2$. Inset shows magnified view of anomaly in magnetization with field cooled (FC) during cooling (FCC) and warming (FCW) cycles respectively. Variation of magnetization as function of magnetic field at selected temperatures is shown in Fig. 1 (b). Inset shows magnified view of the same.

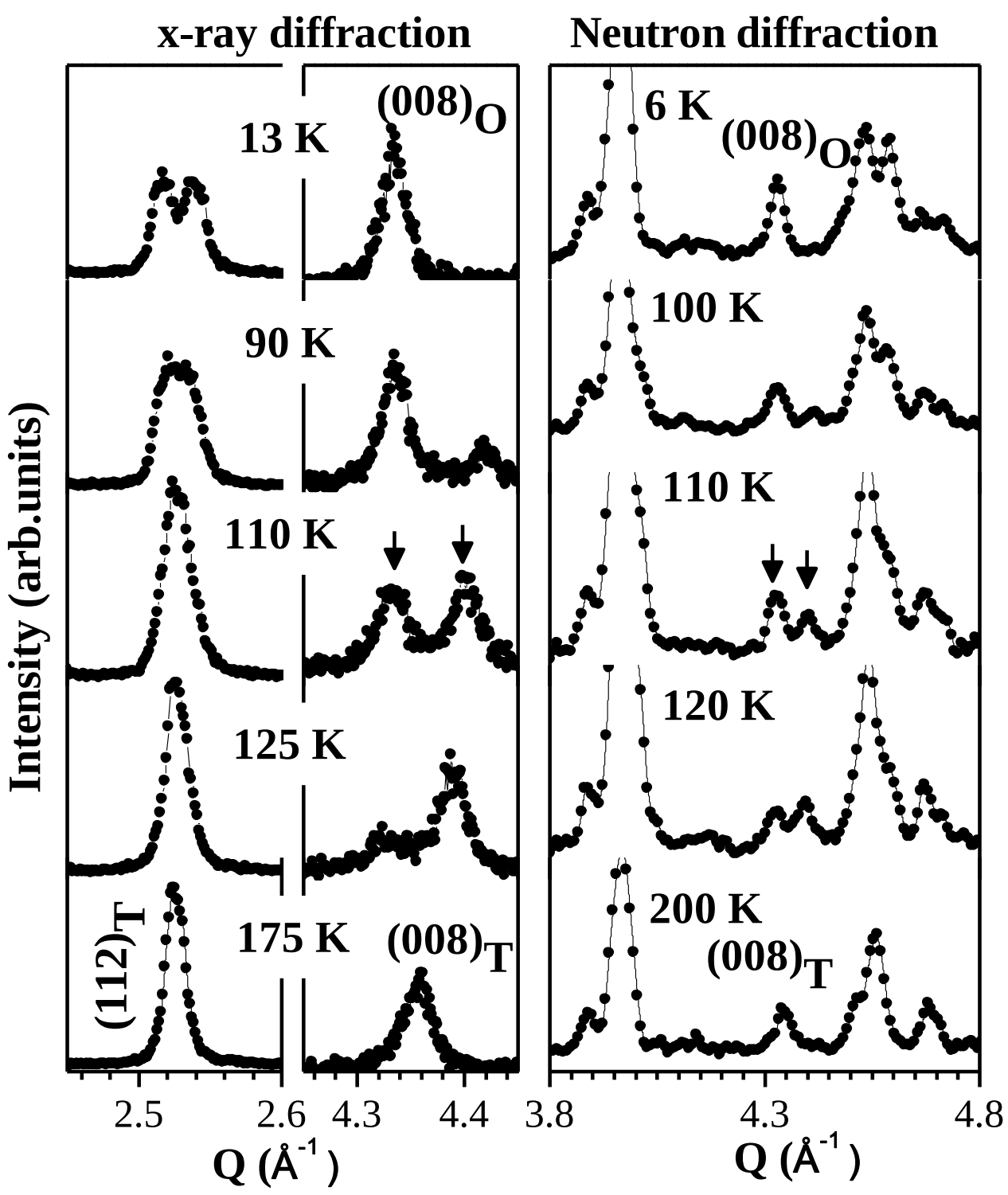

**FIG. 2:** Evolution of the diffraction patterns for $CaFe_2As_2$ at selected temperature. Characteristic reflections of both the phases are marked with arrows. Only relevant part of the patterns is shown for clarity.

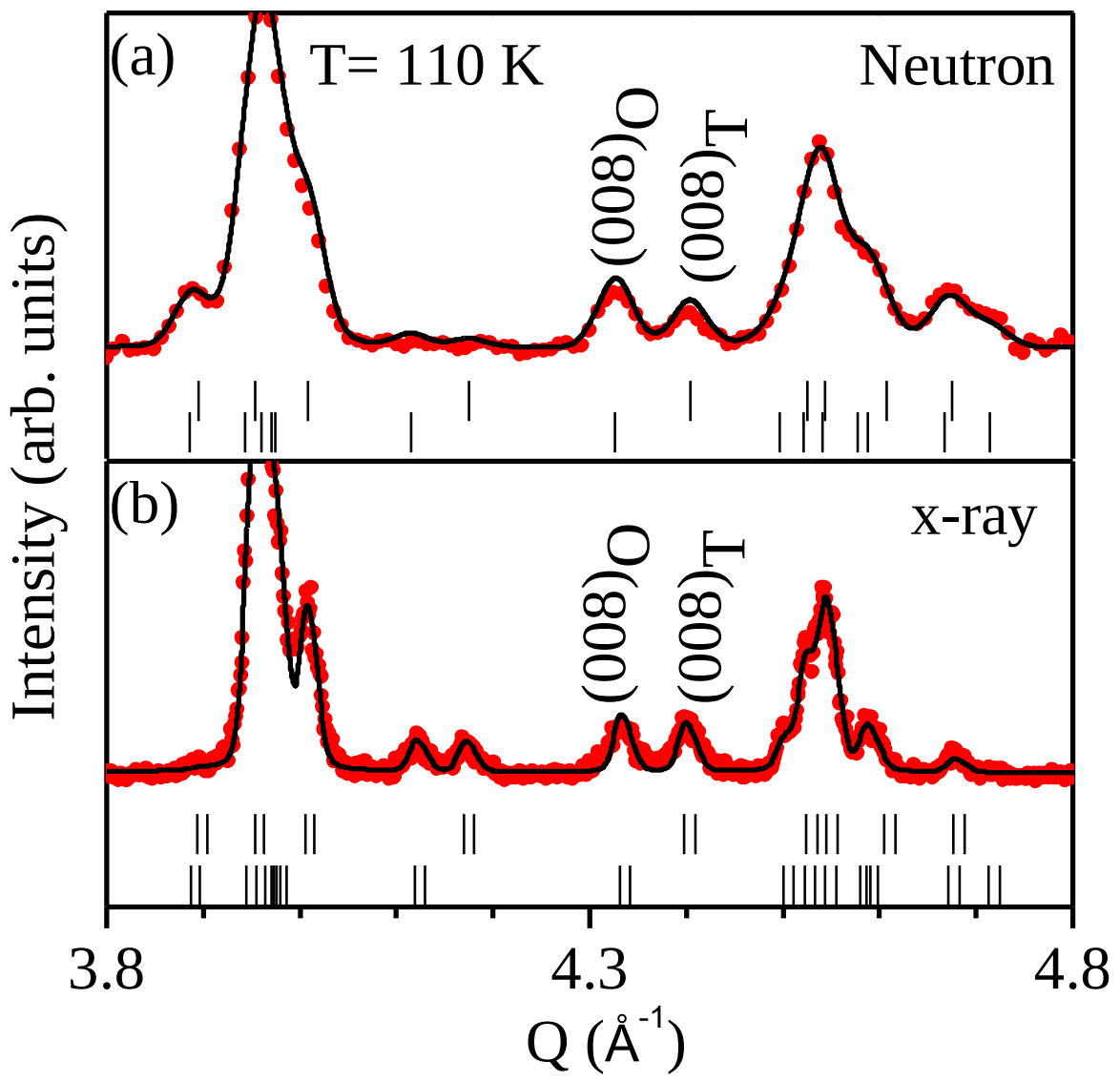

**FIG. 3:** (colour online) Observed (dot), and calculated (continuous line) profiles obtained after the Rietveld refinement of $CaFe_2As_2$ using both orthorhombic (space group *Fmmm*) and tetragonal phases (space group *I4/mmm*) at 110 K.

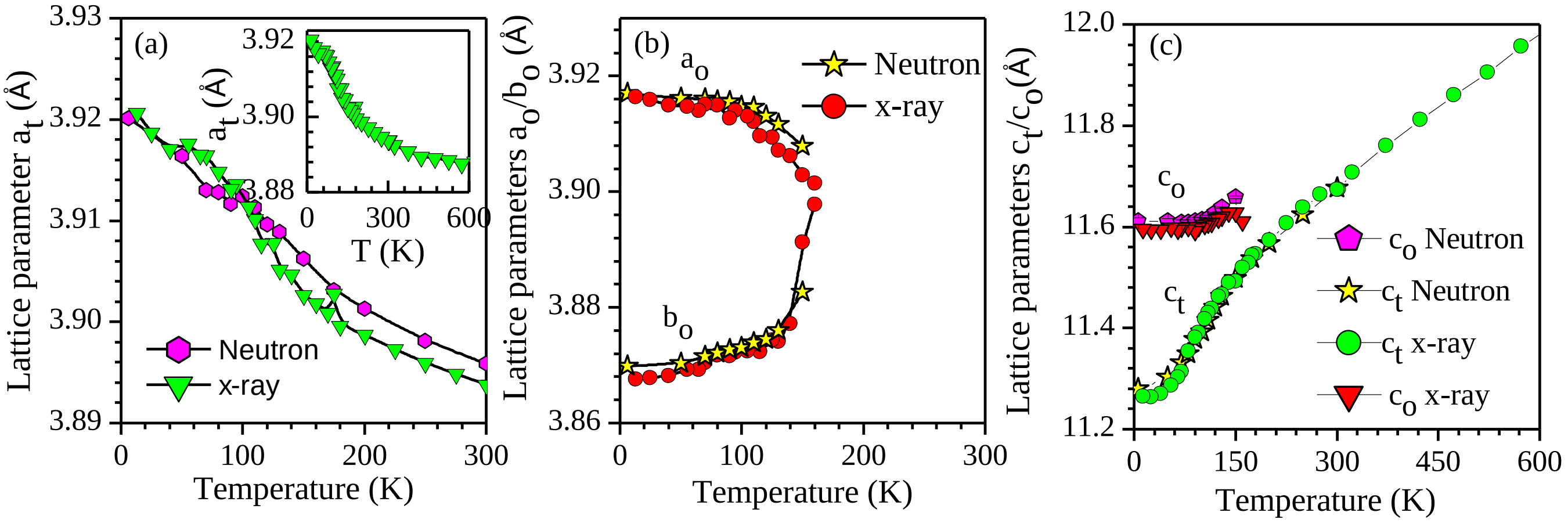

**FIG. 4:** (colour online) Evolution of the obtained lattice parameters from Rietveld refinement for different phases. The subscript t and o refer to the tetragonal and orthorhombic phases respectively in all figures.

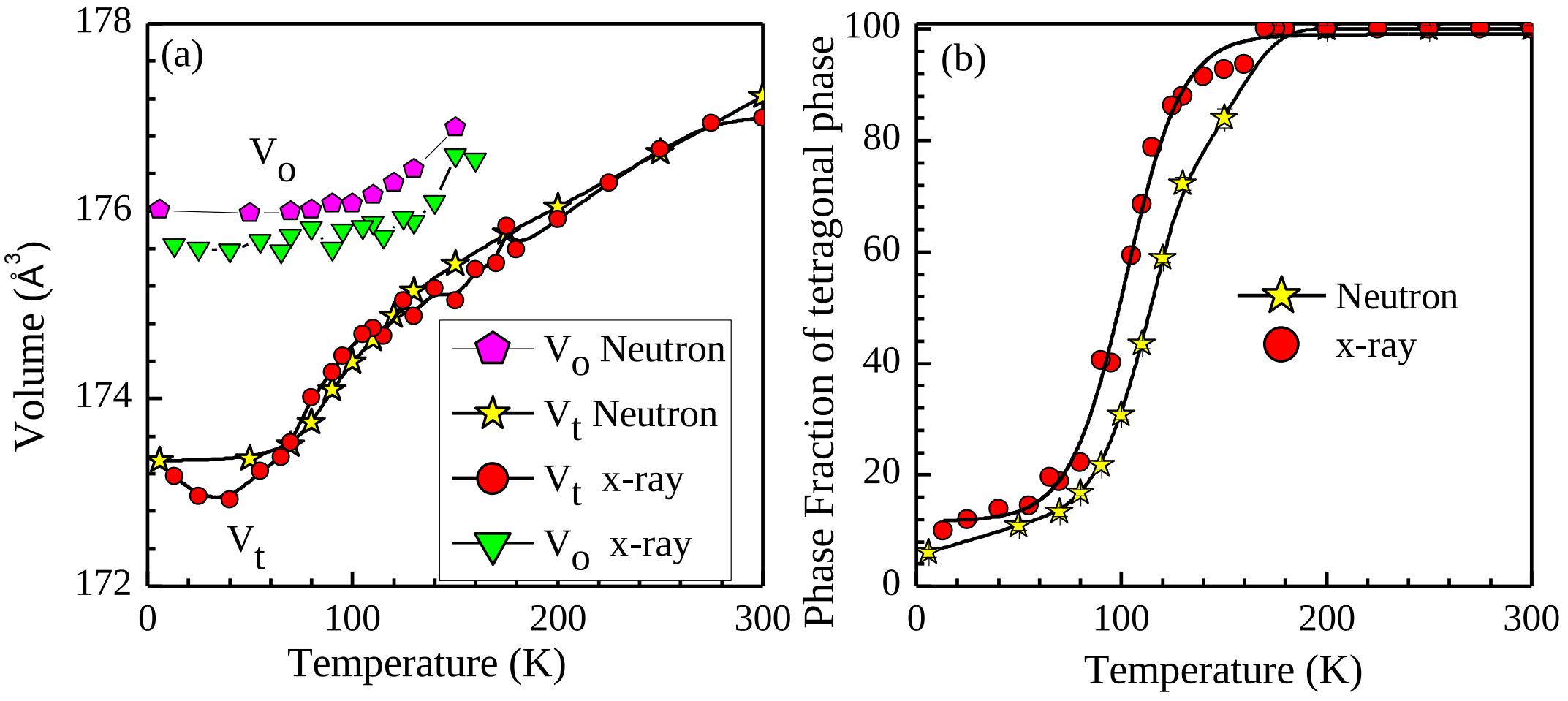

**FIG. 5:** (colour online) Evolution with temperature of the (a) unit cell volume, and (b) phase fraction of the tetragonal phase. For the sake of easy comparison with the tetragonal phase, the unit cell volume of the orthorhombic phase is divided by 2.

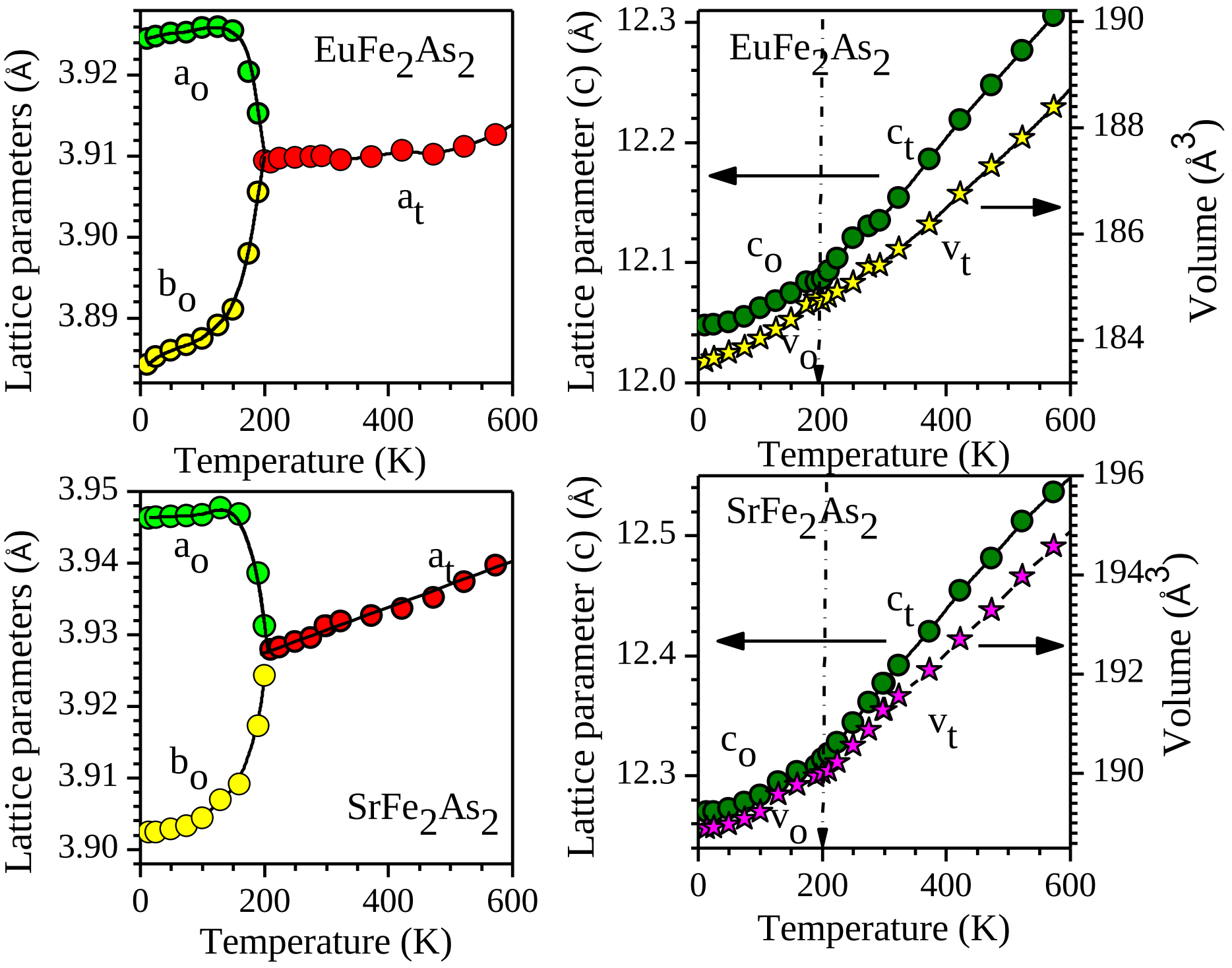

**FIG. 6:** (colour online) Evolution of the structural parameters obtained from Rietveld refinement for $SrFe_2As_2$ and $EuFe_2As_2$ for different phases.

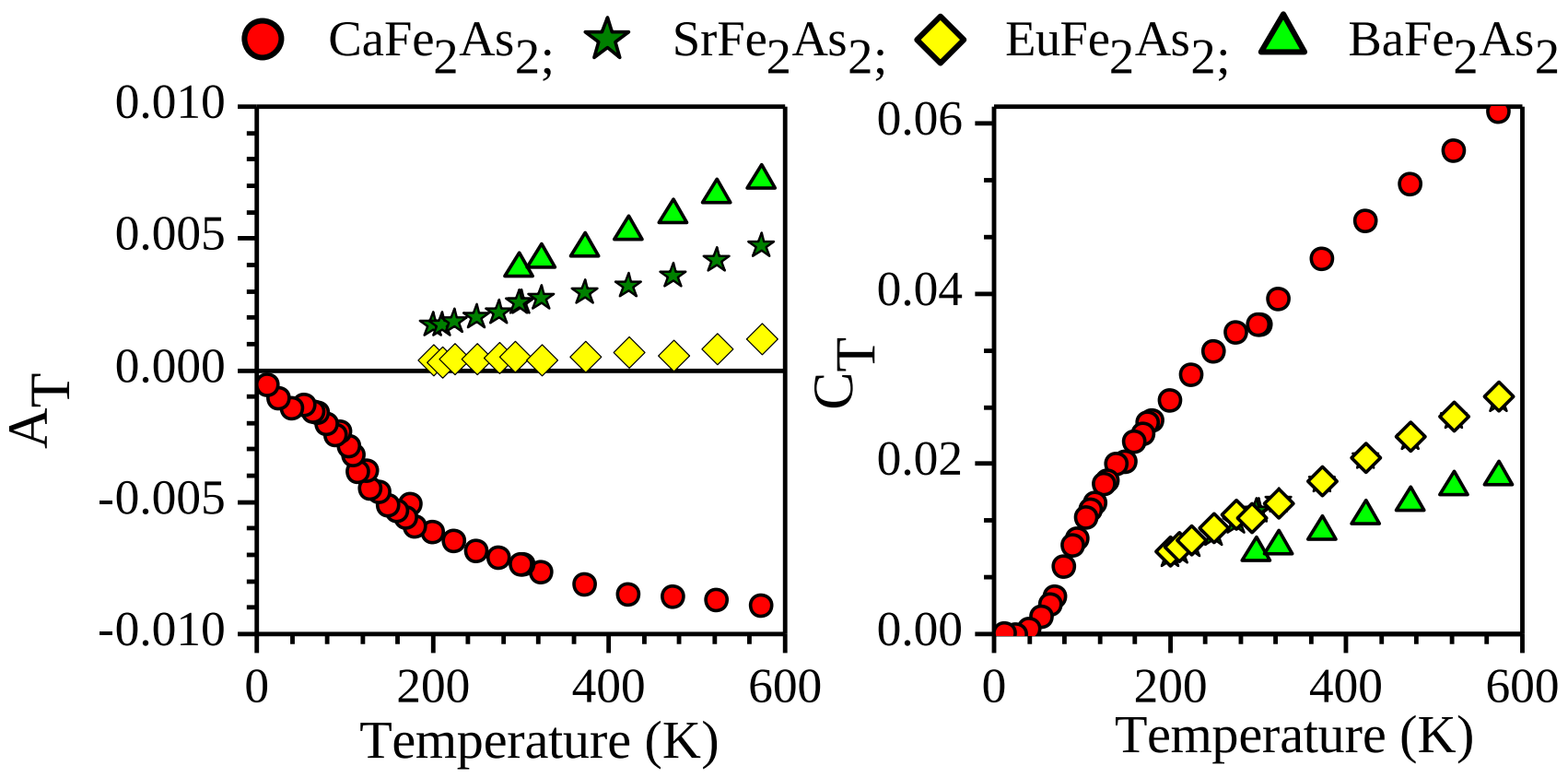

**FIG. 7:** (colour online) Evolution of the observed anisotropic thermal expansivities with temperature for the tetragonal phase.

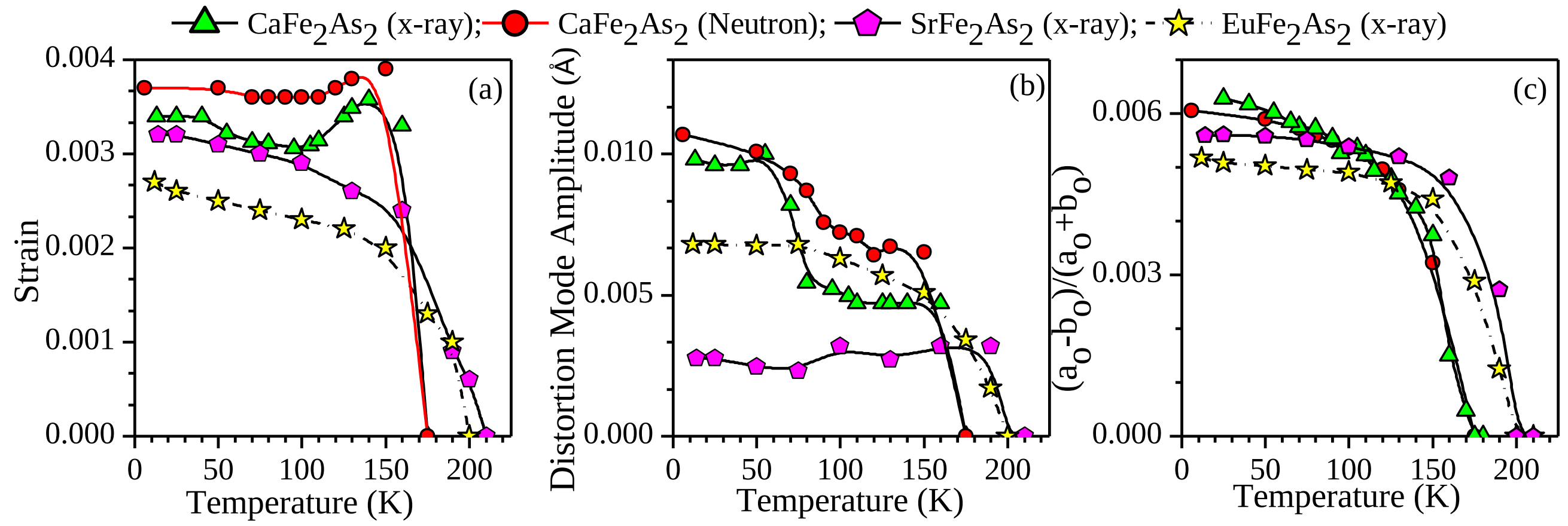

**FIG. 8:** (colour online) Variation of the (a) strain, and (b) amplitude of distortion mode, as calculated using AMPLIMODES for $CaFe_2As_2$, $SrFe_2As_2$ and $EuFe_2As_2$. Temperature dependence of the orthorhombic splitting is shown in (c).